\documentclass[fleqn,usenatbib]{mnras} 

\usepackage[T1]{fontenc}
\usepackage{ae,aecompl}

\usepackage{mathptmx}

\usepackage[dvips]{graphicx}
\usepackage{hyperref}
\usepackage{amsmath}                 
\usepackage{amssymb}                 

\def\bea{\begin{eqnarray}}
\def\eea{\end{eqnarray}}
\def\be{\begin{equation}}
\def\ee{\end{equation}}

\def\de{\Delta}
\def\ms{M_\odot}

\title[Cassiopeia~A and direct URCA cooling]
{Cassiopeia~A and direct Urca cooling}

\author
[G. Taranto, G. F. Burgio, and H.-J. Schulze]
{
G. Taranto,$^{1,2}$
G. F. Burgio,$^1$
and H.-J. Schulze$^1$
\\
$^1$INFN Sezione di Catania, Via Santa Sofia 64, 95123 Catania, Italy \\
$^2$Dipartimento di Fisica e Astronomia, Universit\`a di Catania,
Via Santa Sofia 64, 95123 Catania, Italy
}

\date{\today}
\pubyear{2015}

\begin{document}
\label{firstpage}
\pagerange{\pageref{firstpage}--\pageref{lastpage}}
\maketitle

\begin{abstract}
We model neutron star cooling,
in particular the current rapid cooldown of the neutron star Cas~A,
with a microscopic nuclear equation of state
featuring strong direct Urca processes
and using compatible nuclear pairing gaps as well as effective masses.
Several scenarios are possible to explain the features of Cas~A,
but only large and extended proton $^1S_0$ gaps and small neutron $^3PF_2$ gaps
are able to accommodate also the major part of the complete current cooling data.
We conclude that the possibility of strong direct Urca processes
cannot be excluded from the cooling analysis.
\end{abstract}



\section{Introduction}

With the commissioning of increasingly sophisticated instruments,
more and more details of the very faint signals emitted by neutron stars (NS)
can be quantitatively monitored.
This will allow in the near future an ever increasing accuracy to constrain
the theoretical ideas for the ultra-dense matter that composes these objects.

One important tool of analysis is the temperature-vs.-age cooling diagram,
in which currently a few ($\sim20$) observed NS are located.
NS cooling is over a vast domain of time
($10^{-10}-10^5\,\text{yrs}$)
dominated by neutrino emission due to several microscopic processes
\citep{2001rep,2006ARNPS,2006PaWe,2007LatPra}.
The theoretical analysis of these reactions requires the knowledge of the
elementary matrix elements,
the relevant beta-stable nuclear equation of state (EOS), and, most important,
the superfluid properties of the stellar matter, i.e.,
the gaps and critical temperatures in the different pairing channels.

Even assuming (without proper justification)
the absence of exotic components like hyperons and/or quark matter,
the great variety of required input information under extreme conditions,
that is theoretically not well under or out of control,
renders the task of providing reliable and quantitative predictions
currently extremely difficult.

Recently this activity has been spurred
by the observation of very rapid cooling of the supernova remnant Cas~A,
of current age 335 years and
surface temperature $T \approx 2\times 10^6\,\text{K}$,
for which different analyses deduce a temperature decline of about
2 to 5 percent during the last ten years \citep{2009Nat,2010HeiHo,2013Elsha}.
Mass and radius of this object are not directly observed,
but in recent works optimal values
$M=1.62\,\ms$, $R\approx 10.2\,\text{km}$ \citep{2013Elsha}
or a range
$M=(1.1-1.7)\,\ms$, $R\approx (11.4-12.6)\,\text{km}$ \citep{2015HoPRC}
are reported,
dependent on the assumed EOS.

Two major theoretical scenarios have been proposed to explain this observation:
One is to assume a fine-tuned small neutron 3PF2 (n3P2) gap,
$T_c\approx (5-9)\times10^8\,\text{K}
\sim{\cal O}(0.1\,\text{MeV}$)
\citep{2011PRLPage,2011MNRASYA,2011MNRASSHTE},
that generates strong cooling at the right moment due to the
superfluid neutron pair breaking and formation (PBF) mechanism \citep{pbf};
the other one is based on a strongly reduced thermal conductivity of
the stellar matter that delays the heat propagation
from the core to the crust to a time compatible with the age
of Cas~A \citep{2012Bla,2013Bla}.
Both explanations have in common that they exclude the possibility
of large ($\gtrsim 0.1\,\text{MeV}$) n3P2 gaps;
in the first case because the corresponding critical temperature of the
PBF process has to match the current internal temperature of Cas~A;
in the second case because such a gap would
block too strongly the modified Urca (MU) cooling of the star
and therefore lead to a too high temperature of Cas~A.

Some alternative scenarios have also been brought forward.
Amongst them,
it was suggested in \citep{Bonanno14} that the fast cooling regime
observed in Cas A can be explained if the Joule heating produced by dissipation
of the small-scale magnetic field in the crust is taken into account.
A further explanation was proposed in \citep{Armen13},
according to which the enhancement of the neutrino emission is triggered by
a transition from a fully gapped two-flavor color-superconducting phase
to a gapless/crystalline phase,
although such a scenario requires a very massive $\sim2\ms$ star.

A common feature of all these cooling scenarios is that they exclude
from the beginning the possibility of very fast direct Urca (DU) cooling,
although many microscopic nuclear EOS do reach easily the required
proton fractions for this process \citep{zhl1,2010bs,zhl2,sel};
and we will employ in this work an EOS that does so.
However,
the Akmal-Pandharipande-Ravenhall (APR) variational EOS \citep{1998APR},
which is perhaps the most frequently used EOS for cooling simulations
\citep{2005Gusa,2011PRLPage,2011MNRASYA,2011MNRASSHTE,2012Bla,2013Bla},
(in spite of the fact that it does not reproduce the empirical saturation
point of nuclear matter without an ad-hoc correction),
features a rather low proton fraction and DU cooling only
sets in for very heavy neutron stars, $M\gtrsim2\,\ms$.
Since in any case neither this nor any other EOS can currently be
experimentally verified or falsified at high density,
the frequent use of one particular EOS represents an important bias
that should not be underestimated.

Another critical point of most current cooling simulations
is the fact that EOS and pairing gaps are
treated in disjoint and inconsistent manner, i.e.,
a given EOS is combined with pairing gaps obtained
within a different theoretical approach
and using different input interactions.

In this work we try to improve on both aspects, i.e.,
we include the DU cooling process predicted by our
microscopic nuclear EOS,
and we use compatible nuclear pairing gaps obtained with
exactly the same nuclear (in-medium) interaction.
Furthermore, we also employ recent results for
nucleon effective masses obtained
in the same approach with the same interactions \citep{meff},
which affect the microscopic cooling reactions.

This paper is organized as follows.
In Section~\ref{s:eos} we give a brief overview of the
Brueckner-Hartree-Fock (BHF) theoretical framework adopted for the EOS,
whereas in Section~\ref{s:gap} pairing gaps obtained in the same framework
and with the same interaction, will be introduced.
Section~\ref{s:cool} is devoted to the discussion of several scenarios
for Cas A cooling,
taking into account different choices for superfluid gaps and
thermal conductivity.
Conclusions are drawn in Section~\ref{s:end}.

\section{Equation of state}
\label{s:eos}

Our EOS is determined within the BHF theoretical
approach for nuclear matter \citep{1976Jeu,1999Book,2012Rep},
which computes the in-medium $G$-matrix nucleon-nucleon (NN) interaction
from the bare NN potential $V$,
\be
 G[\rho;\omega] = V + \sum_{k_a k_b} V {{|k_a k_b\rangle Q \langle k_a k_b|}
 \over {\omega - e(k_a) - e(k_b)}} G[\rho;\omega] \:,
\label{e:g}
\ee
where
$\rho=\sum_{k<k_F}$ is the nucleon number density,
and $\omega$ the starting energy. The single-particle (s.p.) energy
\be
 e(k) = e(k;\rho) = {k^2\over 2m} + U(k;\rho)
\label{e:en}
\ee
and the Pauli operator $Q$ determine the propagation
of intermediate baryon pairs.
The BHF approximation for the s.p.~potential
$U(k;\rho)$ using the {\em continuous choice} prescription is
\be
 U(k;\rho) = {\rm Re} \sum_{k'<k_F}
 \big\langle k k'\big| G[\rho; e(k)+e(k')] \big| k k'\big\rangle_a \:,
\ee
where the subscript $a$ indicates antisymmetrization of the matrix element,
and the energy per nucleon is then given by
\be
 {E \over A} =
 {3\over5}{k_F^2\over 2m} + {1\over{2\rho}} \sum_{k<k_F} U(k;\rho)
\:.
\ee
In this scheme, the only input quantity needed is the bare NN interaction
$V$ in the Bethe-Goldstone equation~(\ref{e:g}),
supplemented by a suitable three-nucleon force (TBF) in order to reproduce
correctly the saturation properties of nuclear matter.
In this work we use the Argonne $V_{18}$ NN interaction \citep{v18}
and the Urbana-type UIX TBF
\citep{Carlson:1983kq,Schiavilla:1985gb,1997Pud}
as input.
The results for the EOS including numerical parametrizations can be found in
\citep{2010bs,sel}.
We also reiterate that in the cooling simulations we employ neutron and proton
effective masses,
\be
 \frac{m^*(k)}{m} = \frac{k}{m} \left[ \frac{d e(k)}{dk} \right]^{-1} \:,
\ee
derived consistently from the BHF s.p.~energy $e(k)$, Eq.~(\ref{e:en}),
\citep{meff}.
Although the effect is not large compared to other uncertainties regarding
the cooling,
such a consistent treatment is hard to find in previous works.
The model just described will be denoted by ``BHF" in the following
and some confrontation with the ``APR" model \citep{1998APR},
which is based on the same input interactions,
will be made.
We use here the original APR
``$A18 + \delta v +\text{UIX}^*~\text{corrected}$''
results \citep{1998APR}
and not one of the parametrized versions \citep{hhj,2005Gusa},
where the high-density behavior is arbitrarily modified.

\begin{figure}
\includegraphics[angle=0,scale=0.74,clip]{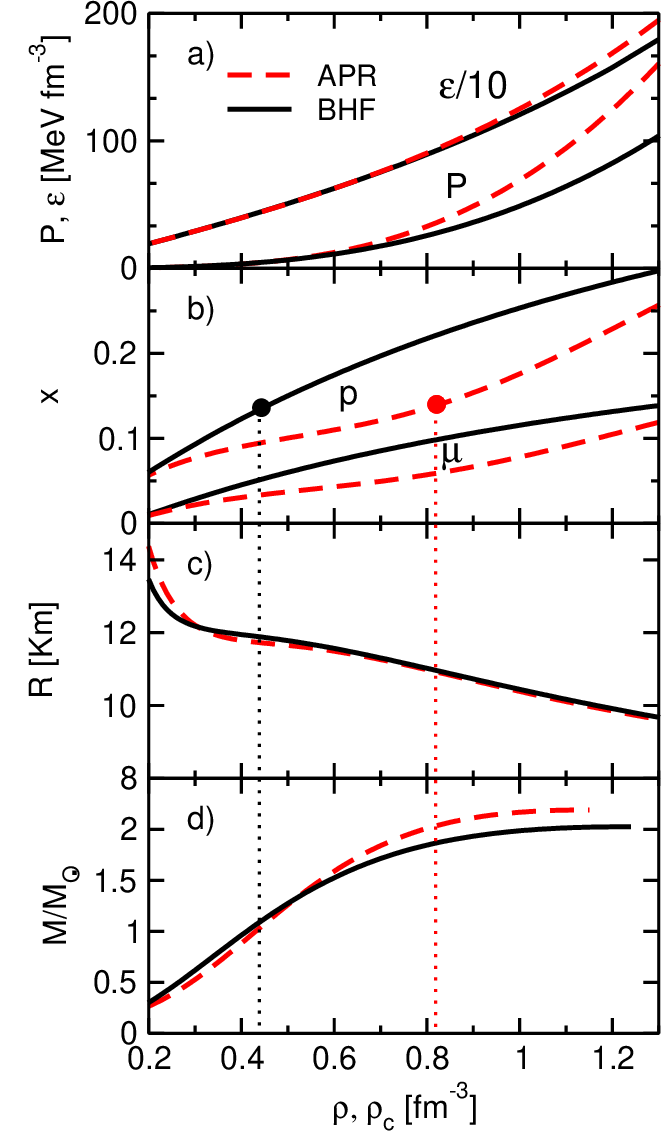}
\caption{(Color online)
Pressure and energy density (a),
and proton and muon fractions (b)
as functions of baryon number density $\rho$
in beta-stable matter for the APR and BHF EOS.
The lower panels show
neutron star mass (d) and radius (c)
as functions of the central density $\rho_c$.
The DU onset is indicated by vertical dotted lines.
}
\label{f:eos}
\end{figure}

\begin{figure*}
\includegraphics[angle=0,scale=0.5,clip]{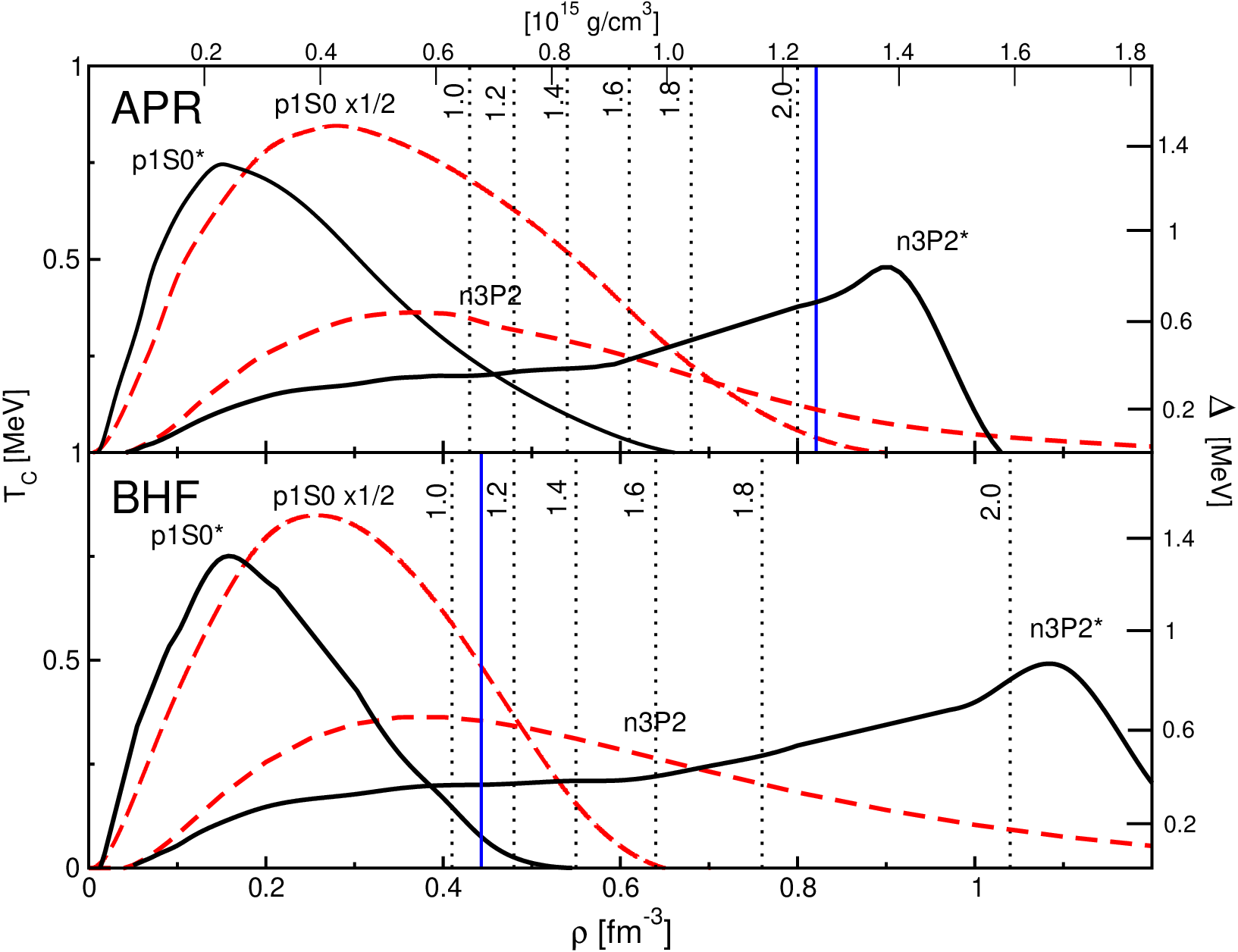}
\caption{(Color online)
Pairing gaps in NS matter
for the APR and BHF models
in the p1S0 and n3P2 channels,
including ($^*$) or not effective mass effects.
Vertical dotted lines indicate the central density
of NS with different masses
$M/\ms=1.0,\ldots,2.0$.
DU onset occurs at the vertical solid (blue) lines.
Note the scaling factor 1/2 for the p1S0 curves.
}
\label{f:gaps}
\end{figure*}

In fact we compare in Fig.~\ref{f:eos}
the NS EOS obtained with the APR and BHF models, i.e.,
panels (a) and (b) show
pressure, energy density, and proton and muon fractions
of beta-stable and charge-neutral matter
as functions of the baryon density.
It is obvious that the crucial difference between both models is the
much higher proton fraction in the BHF approach.
In this case the DU threshold is already reached at
$\rho=0.44\,\text{fm}^{-3}$ ($x_p=0.136$),
whereas with the APR it is delayed to
$\rho=0.82\,\text{fm}^{-3}$ ($x_p=0.140$).

On the contrary, the APR EOS is somewhat stiffer, i.e.,
features a larger pressure and energy density.
We also mention that the APR EOS becomes superluminal at
$\rho=0.85\,\text{fm}^{-3}$,
whereas BHF remains always below the critical threshold \citep{2010bs,sel}.

By solving the standard Tolman-Oppenheimer-Volkov equations
for the NS structure,
this input yields the NS (mass,radius) -- central density
relations shown in panels (c) and (d) of Fig.~\ref{f:eos}.
We remark that both models reach maximum masses (slightly)
above two solar masses and predict very similar radii
in spite of their different matter composition.
With BHF the DU process is active in nearly all stars, $M/\ms>1.10$,
while with APR only in the most heavy ones, $M/\ms>2.03$.
This has profound consequences for the cooling behavior.

\section{Pairing gaps and critical temperatures}
\label{s:gap}

Of vital importance for any cooling simulation is the knowledge of the
1S0 and 3PF2 pairing gaps for neutrons and protons in beta-stable matter,
which on one hand block the DU and MU reactions,
and on the other hand open new cooling channels by the PBF mechanism
for stellar matter
in the vicinity of the critical temperature \citep{2001rep}.
As usual, we focus in this work on the most important proton 1S0 (p1S0)
and neutron 3PF2 (n3P2) pairing channels and neglect the less important
remaining combinations.

As stressed before, the gaps should be computed in a framework that is
consistent with the determination of the EOS, i.e.,
be based on the same NN interaction and using the same medium effects
(TBF and effective masses),
and indeed we follow this procedure here,
by using the results of \citep{ourgaps},
which employed the same V18+UIX nuclear interaction and BHF s.p.~spectra
for the calculation of the gaps.
To be more precise,
and focusing on the more general case of pairing in the coupled 3PF2 channel,
the pairing gaps were computed on the
BCS level by solving the (angle-averaged) gap equation
\citep{bcsp1,bcsp2,bcsp3,bcsp4,bcsp5,bcsp6}
for the two-component $L=1,3$ gap function,
\be
  \left(\!\!\!\begin{array}{l} \de_1 \\ \de_3 \end{array}\!\!\!\right)\!(k) =
  - {1\over\pi} \int_0^{\infty}\!\! dk' {k'}^2 {1\over E(k')}
  \left(\!\!\!\begin{array}{ll}
   V_{11} & V_{13} \\ V_{31} & V_{33}
  \end{array}\!\!\!\right)\!(k,k')
  \left(\!\!\!\begin{array}{l} \de_1 \\ \de_3 \end{array}\!\!\!\right)\!(k')
\label{e:gap}
\ee
with
\be
  E(k)^2 = [e(k)-\mu]^2 + \de_1(k)^2 + \de_3(k)^2 \:,
\ee
while fixing the (neutron or proton) density,
\be
  \rho = {k_F^3\over 3\pi^2}
   = 2 \sum_k {1\over 2} \left[ 1 - { e(k)-\mu \over E(k)} \right] \:.
\label{e:rho}
\ee
Here $e(k)$ are the BHF s.p.~energies, Eq.~(\ref{e:en}),
containing contributions due to two-body and three-body forces,
$\mu \approx e(k_F)$ is the (neutron) chemical potential
determined self-consistently from Eqs.~(\ref{e:gap}--\ref{e:rho}),
and
\be
   V^{}_{LL'}(k,k') =
   \int_0^\infty \! dr\, r^2\, j_{L'}(k'r)\, V^{TS}_{LL'}(r)\, j_L(kr)
\label{e:v}
\ee
are the relevant potential matrix elements
($T=1$ and
$S=1$; $L,L'=1,3$ for the 3PF2 channel,
$S=0$; $L,L'=0$ for the 1S0 channel)
with
\be
 V = V_{18} + \bar{V}_\text{UIX} \:,
\label{e:v23}
\ee
composed of two-body force and averaged TBF.

The relation between (angle-averaged) pairing gap at zero temperature
$\de \equiv \sqrt{\de_1^2(k_F)+\de_3^2(k_F)}$
obtained in this way and the critical temperature of superfluidity is then
$T_c \approx 0.567\de$.

Fig.~\ref{f:gaps} displays the p1S0 and n3P2 pairing gaps
as a function of baryonic density of beta-stable matter
for the APR (upper panel) and BHF (lower panel) models.
Also indicated are the central densities of NS with different masses,
in order to easily identify which range of gaps is active in different stars.
Note that the in-medium modification of the pairing interaction
is treated consistently
(via the compatible s.p.~energies and TBF)
only in the BHF model.

In \citep{ourgaps} different levels of approximation
for the calculation of gaps were discussed,
in particular one including only the two-body force $V_{18}$ in Eq.~(\ref{e:v23})
together with the kinetic s.p.~energies,
and another one
(curves denoted by p1S0* and n3P2* in Fig.~\ref{f:gaps})
using the BHF s.p. spectra according to Eq.~(\ref{e:en}).
Note that polarization corrections \citep{pol1,pol2,pol,pol3}
were not considered in that work,
which for the case of the 1S0 channel are known to be repulsive,
but for the 3PF2 are still essentially unknown,
and might change the value of the gaps even by orders of magnitude
\citep{ppol1,ppol2,ppol3}.
In order to represent this uncertainty,
we use in the cooling simulations the density dependence of the
pairing gaps shown in Fig.~\ref{f:gaps},
but employ global scaling factors $s_p$ and $s_n$, respectively.

Qualitatively one observes in Fig.~\ref{f:gaps} the
natural scaling effect of the different proton fractions
for the BHF and APR EOS,
such that the p1S0 gaps extend to larger (central) densities for the APR model,
due to the lower proton fraction in that case.
Therefore the blocking effect on the cooling extends up to higher densities
and NS masses for the APR model.
The crucial difference is again the onset of the DU process,
which is active for nearly all NS in the BHF case.
However, the n3P2$^{(*)}$ gaps extend up to very large density
and can thus provide an efficient means to block this cooling process,
in particular for the n3P2$^*$ model comprising medium effects.
The price to pay is an enhanced PBF cooling rate close to the
critical temperature in that case.
Note that the 3P2 gaps shown in the figure are larger than those
currently employed in cooling simulations,
which will be discussed in the next section,
and that at the moment there exists no satisfactory theoretical
calculation of p-wave pairing that includes consistently all medium effects.

\section{Neutron star cooling and C\uppercase{as}~A}
\label{s:cool}

Having quantitatively specified EOS and pairing gaps,
the NS cooling simulations are carried out using the
widely used code {\tt NSCool} \citep{Pageweb},
which comprises all relevant cooling reactions:
DU, MU, PBF, and Bremsstrahlung.

A further important ingredient of the simulations is the
(leptonic and baryonic) thermal conductivity,
and the default choice in \citep{2006ARNPS,2006PaWe,Pageweb}
is to use the results of \citep{gnedin,2001rep,baiko}.
Recently it has been conjectured
\citep{shternin1,shternin2,2012Bla,2013Bla}
that the conductivities could be substantially
(by one order of magnitude) suppressed by in-medium effects,
and this has been put forward as an alternative explanation of the rapid Cas~A
cooling.
We follow this idea by introducing a further global scale factor $s_\kappa$
multiplying the total thermal conductivity.
Therefore our calculations are controlled by the three global parameters
$s_p,s_n,s_\kappa$,
and we present now some selected results for certain parameter choices.

\begin{figure}
\includegraphics[angle=0,scale=0.35,clip]{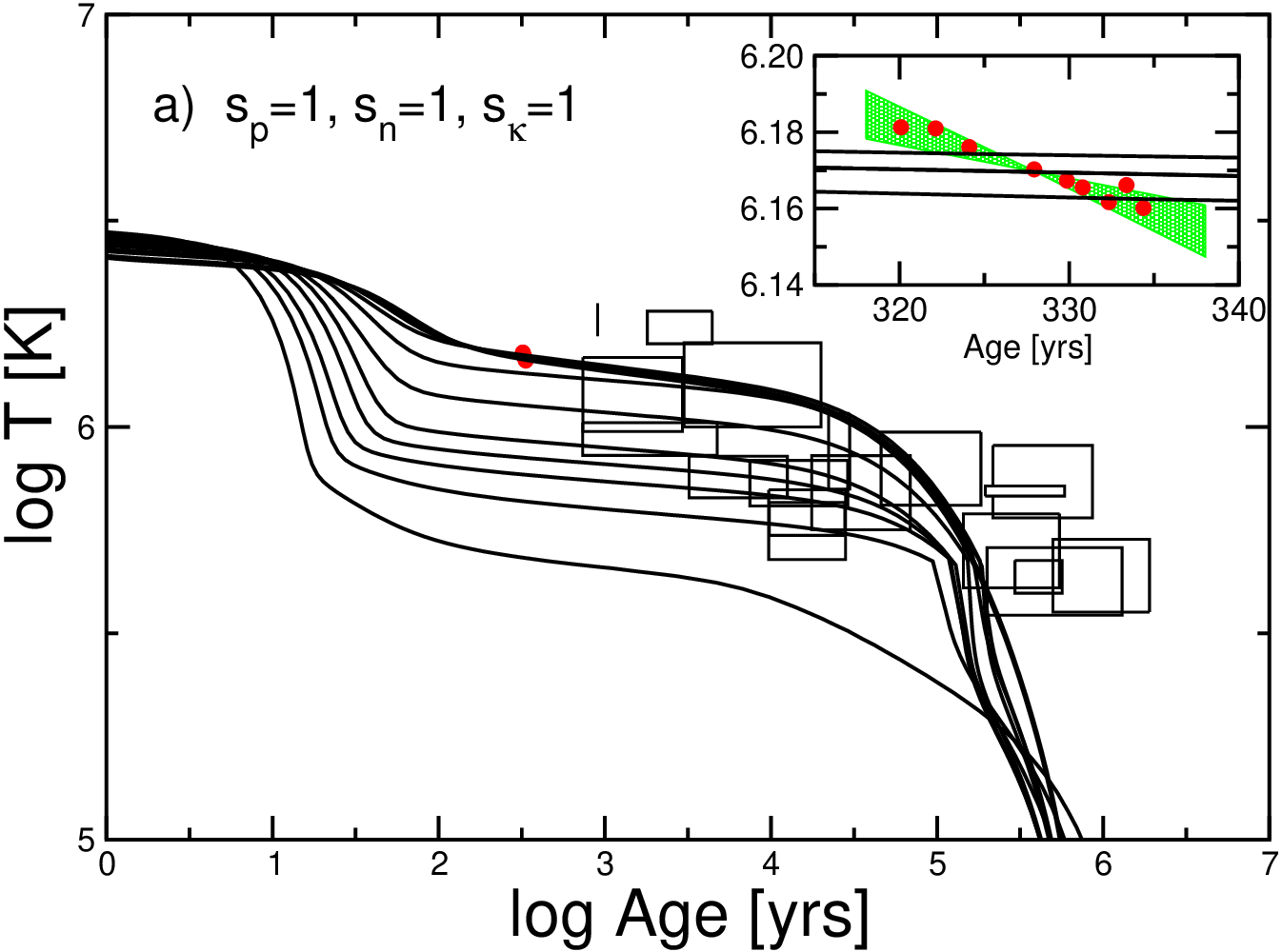}
\includegraphics[angle=0,scale=0.35,clip]{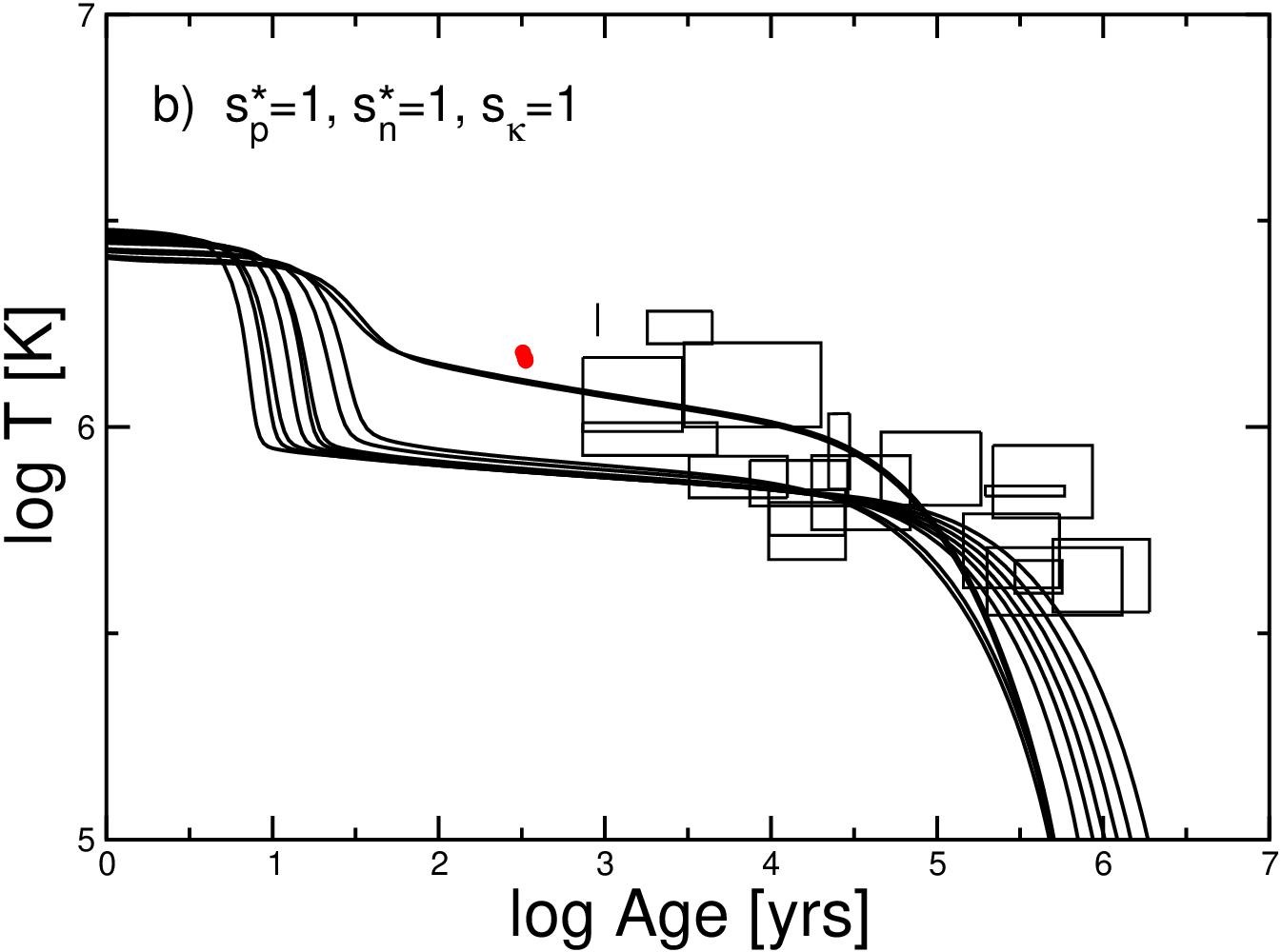}
\caption{(Color online)
Cooling curves with the BHF EOS and no scaling factors,
for different NS masses $M/\ms=1.0,1.1,\ldots,2.0$
(decreasing curves).
The upper panel employs BCS gaps with free s.p.~spectra,
whereas in the lower panel the BHF s.p.~spectra are used.
Red dots show the Cas~A cooling data
(enlarged in the inset of the upper panel;
the $M/\ms=$1.1 and 1.2 curves overlap).
}
\label{f:bcs}
\end{figure}

Our set of observational cooling data comprises the
(age, temperature) information
of the 19 isolated NS sources listed in \citep{Beznogov1}.
We point out, however, the recent discovery of the unusually hot object
XMMU J173203.3-344518 \citep{Klochkov}
with estimated age and temperature
(10-40 kyr, 2.1-2.8 MK).
We do not include it in our current analysis,
as it is in fact incompatible with most previous cooling simulations,
and will certainly be studied in great detail in the future.

\subsection{Scenario 1: Original model, no scaling}

For the sake of illustration and better understanding we begin by showing the
results obtained with the original pairing gaps shown in Fig.~\ref{f:gaps},
and without any modification of the conductivities,
i.e., setting $s_p=s_n=s_k=1$.
Moreover we use the neutron $^1S_0$ BCS gap as calculated in \citep{ourgaps},
without any rescaling,
although the beta-stable matter in the relevant subnuclear density domain
of the crust is inhomogeneous and therefore more elaborate considerations
should be done \citep{inhomo1,inhomo2}.

The upper panel of Fig.~\ref{f:bcs} shows the temperature vs age results
(11 curves for NS with masses $1.0,1.1,\ldots,2.0$)
obtained with the BCS gaps without any medium modification
(dashed red curves in the lower panel of Fig.~\ref{f:gaps}),
while the lower panel employs the gaps with BHF effective masses
(black curves in Fig.~\ref{f:gaps}),
which is indicated by the notation
$s_p^*=s^*_n=1$
here and in the following.
In all cases a heavy (Fe) elements atmosphere ($\eta=0$) has been assumed.
One observes results in line with the features of the pairing gaps,
namely in the upper panel light ($M\lesssim1.4\,\ms$) NS cool slower
and heavy ($M\gtrsim1.7\,\ms$) NS cool faster than in the lower panel.
This is due to the larger overall values of the corresponding BCS gaps
in the low-density ($n\lesssim0.6\,\text{fm}^{-3}$) region
and the smaller n3P2 value in the high-density ($n\gtrsim0.7\,\text{fm}^{-3}$)
domain,
see Fig.~\ref{f:gaps},
which cause, respectively,
a stronger or weaker blocking of the dominant DU process
in light or heavy stars.

Very old and warm stars (PSR~B1055-52, RX~J0720.4-3125)
(as well as the recent XMMU J1732)
are not covered by any cooling curve,
just as in the equivalent investigation within the APR model of \citep{pbf};
and we refer to that article for a discussion of possible reasons.
Altogether, our cooling curves for warm stars appear quite similar
to those in that reference,
while there is no difficulty at all to explain cold stars due
to the DU mechanism in the BHF model.
The main reason for the too low temperature of old stars
is an early cooldown due to the n3P2 PBF process,
as we shall see.

The major shortcoming of both scenarios
in Fig.~\ref{f:bcs}
is that they cannot reproduce the
particular cooling properties of Cas~A:
While the first one can fit its current age and temperature
as a $M=1.2\,\ms$ NS,
neither reproduces the apparent very fast current cooldown
\citep{2009Nat,2010HeiHo,2013Elsha},
shown in the inset of the upper panel.
Precisely for this reason special scenarios with fine-tuned parameters
have been developed, which we analyze now.

\begin{figure}
\includegraphics[angle=0,scale=0.35,clip]{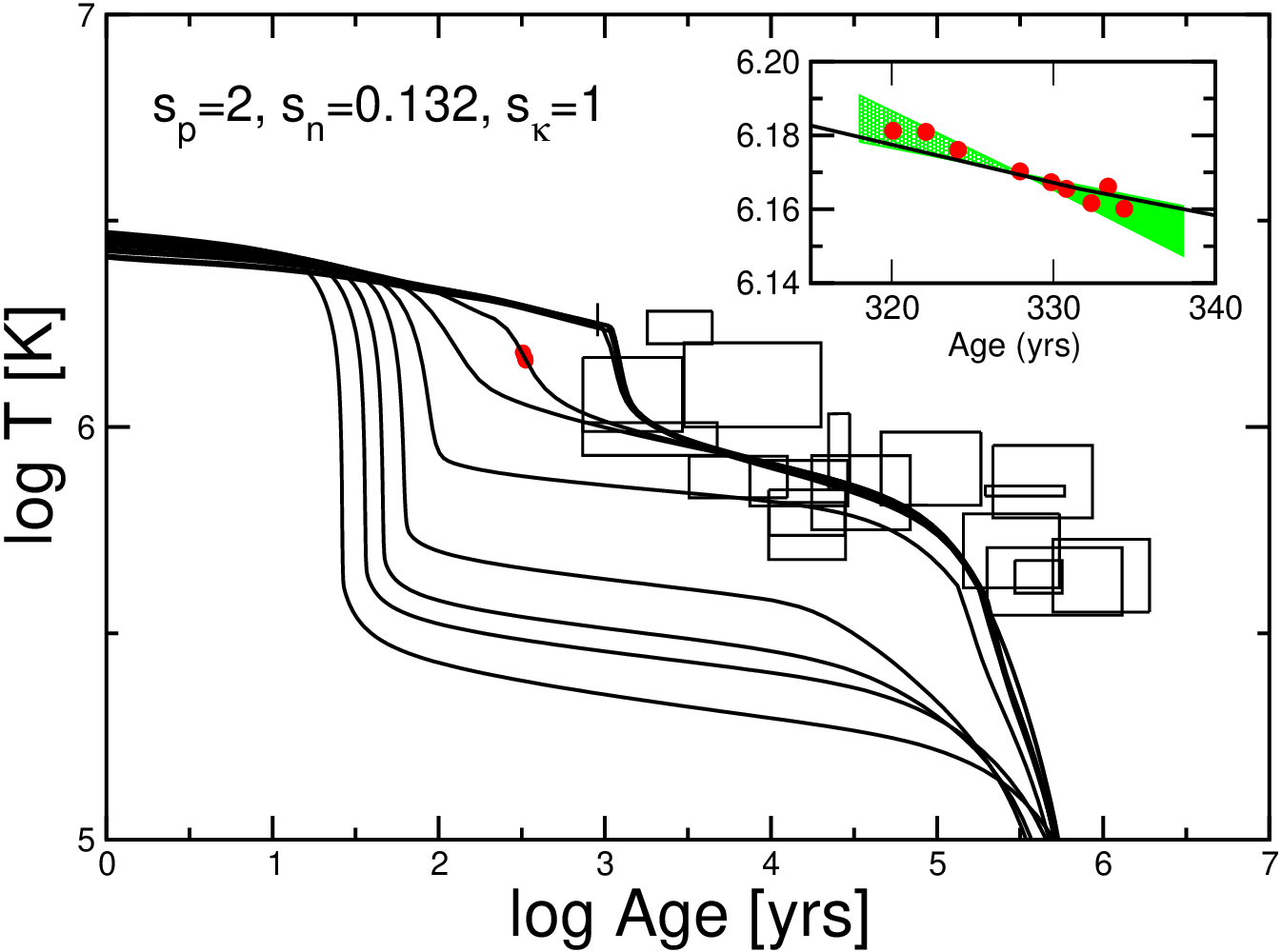}
\caption{(Color online)
Same as Fig.~\ref{f:bcs},
for the PBF cooling scenario.
The $M=1.4\,\ms$ cooling curve passes through Cas~A by construction.
}
\label{f:npb}
\end{figure}

\subsection{Scenario 2: Neutron pair breaking cooling}

A frequent explanation of the rapid cooling of Cas~A is
the one based on an appropriately chosen n3P2 gap,
which causes strong cooling
due to the opening of the neutron PBF process
at the current age/temperature of the star.
The BHF EOS including strong DU reactions
also allows this interpretation by choosing the scaling factors
$s_p=2.0$, $s_n=0.132$, $s_\kappa=1$
for a $1.4\,\ms$ star,
corresponding to maximum values of the gaps
$\de_p\approx 6\,\text{MeV}$ and
$\de_n\approx 0.1\,\text{MeV}$, i.e.,
$\de_p$ is larger than usually chosen
(in order to block the fast DU reaction;
the domain of the p1S0 gap fully covers NS up to about $1.6\,\ms$,
see Fig.~\ref{f:gaps}),
while $\de_n$ is in line with the equivalent results of
\citep{2015HoPRC,2011PRLPage,2011MNRASYA,2011MNRASSHTE}.

The results are shown in Fig.~\ref{f:npb},
which displays the sequence of cooling curves for NS masses
$M/\ms=1.0,1.1,\ldots,2.0$,
the 1.4 case corresponding to Cas~A by construction;
(this might be changed within reasonable limits by choosing different
scaling factors).
In this case the neutron $^1S_0$ gap has been rescaled by a factor 0.04
for finetuning.
As in similar investigations \citep{2015HoPRC,pbf},
one notes that the rapid cooldown caused by the n3P2 PBF
renders even more difficult the reproduction of old hot stars.
Also for this reason alternative scenarios have been developed,
and we analyze one of them now.

\begin{figure}
\includegraphics[angle=0,scale=0.35,clip]{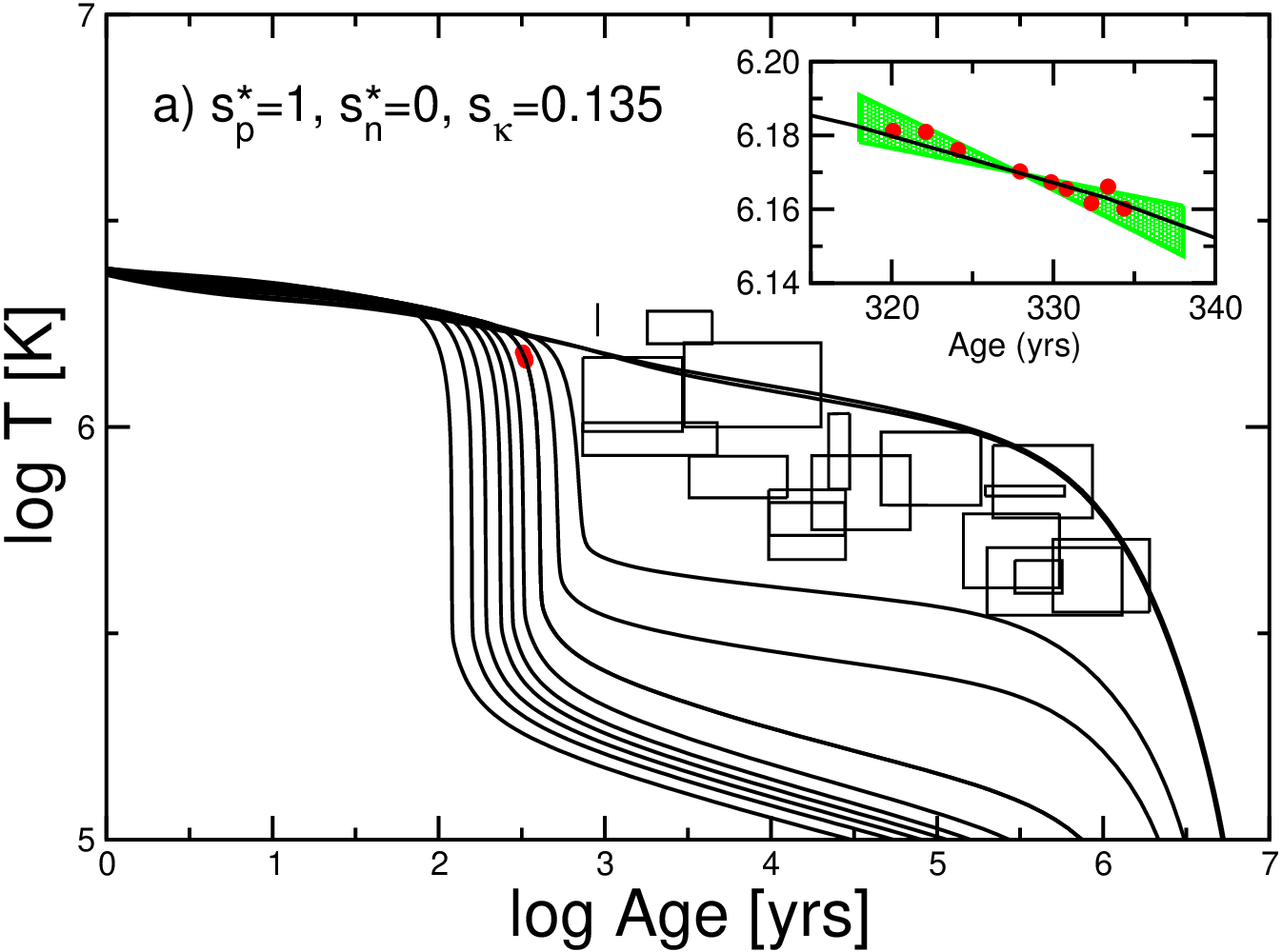}
\includegraphics[angle=0,scale=0.35,clip]{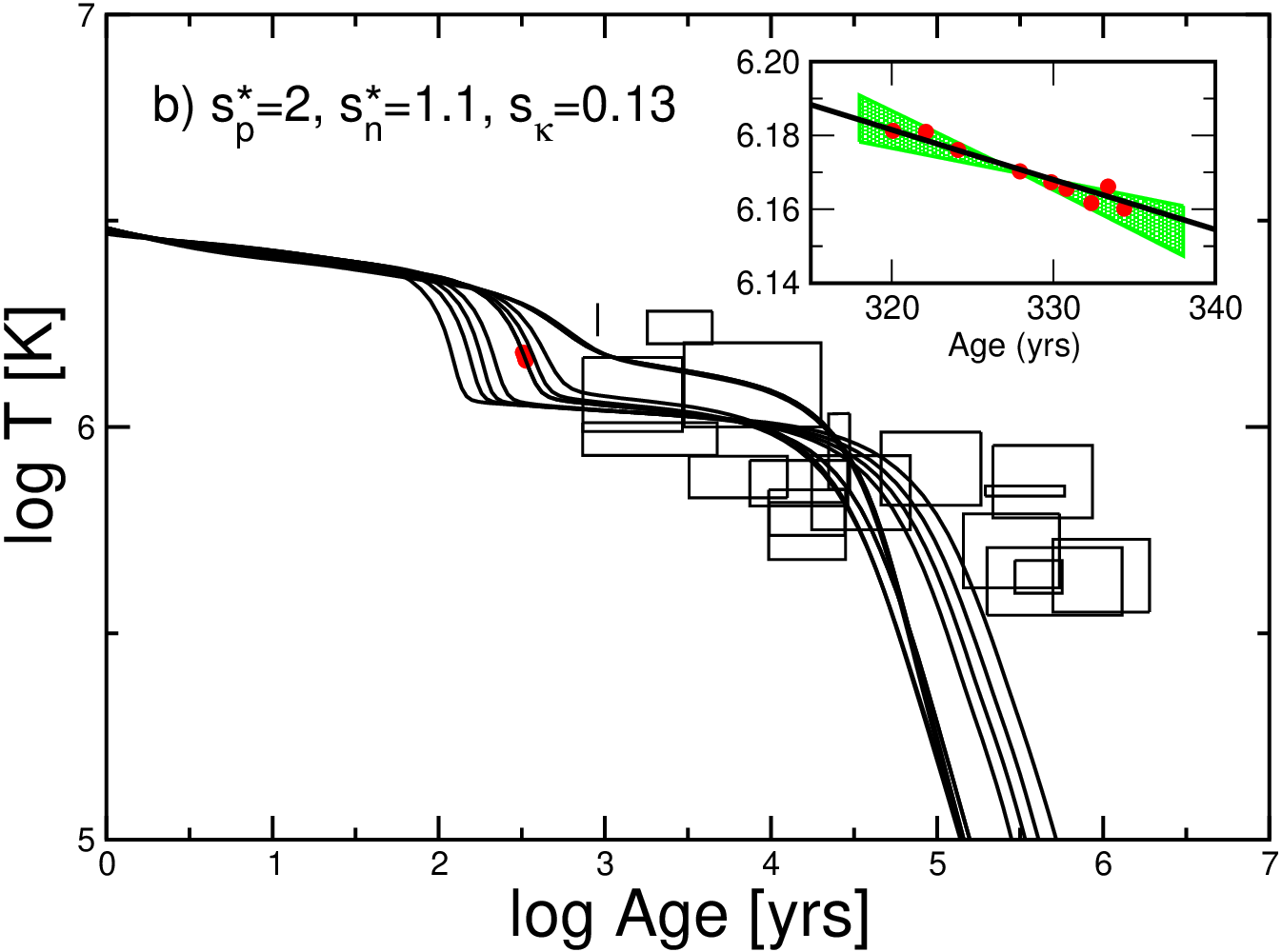}
\caption{(Color online)
Same as Fig.~\ref{f:npb},
for two delayed cooling scenarios.
}
\label{f:del}
\end{figure}

\subsection{Scenario 3: Suppressed thermal conductivity}

The approach of \citep{2012Bla,2013Bla}
features strongly suppressed lepton and baryon thermal conductivities,
which we roughly simulate by the scaling parameter $s_\kappa$,
as in \citep{2012Bla}.
[In \citep{2013Bla} a more microscopic treatment of this reduction
was introduced, which however
did not lead to qualitatively different conclusions].
The reduced conductivity
serves to delay the temperature decline up to the current age of
Cas~A without need to introduce fine-tuned nPBF cooling.
A further peculiarity of this model is the fact that the standard MU
cooling is strongly enhanced by assumed in-medium effects (MMU),
which provides fast cooling for heavy NS,
without need of DU cooling \citep{GriVos}.

Our EOS including DU cooling
also accommodates the possibility
of reduced thermal conductivities,
as demonstrated in Fig.~\ref{f:del}.
The upper panel shows a rather satisfactory fit of all cooling data
including Cas~A,
employing the parameter set
$s_p^*=1$, $s_n^*=0$, $s_\kappa=0.135$,
where the size of $s_\kappa$ is comparable
to the values of about 0.2 deduced in \citep{2012Bla,2013Bla}.

In this scenario a rather small value of the n3P2 gap seems to be required,
as otherwise old hot (and also young cold) NS cannot be obtained,
even if Cas~A is reproduced.
This is demonstrated by a typical result
($s_p^*=2.0$, $s_n^*=1.1$, $s_\kappa=0.13$,
the n1S0 gap has been reduced by 0.09, and $\eta=0.03$ here),
shown in the lower panel of the figure,
where large values for both gaps are used.
The features of Cas~A are reproduced correctly,
but the finite large n3P2 gap causes an early rapid cooldown
incompatible with the temperature of most old NS,
but at the same time together with the large p1S0 gap reduces too
strongly the DU and MU cooling in order to fit young cold stars.

\begin{figure}
\includegraphics[angle=0,scale=0.35,clip]{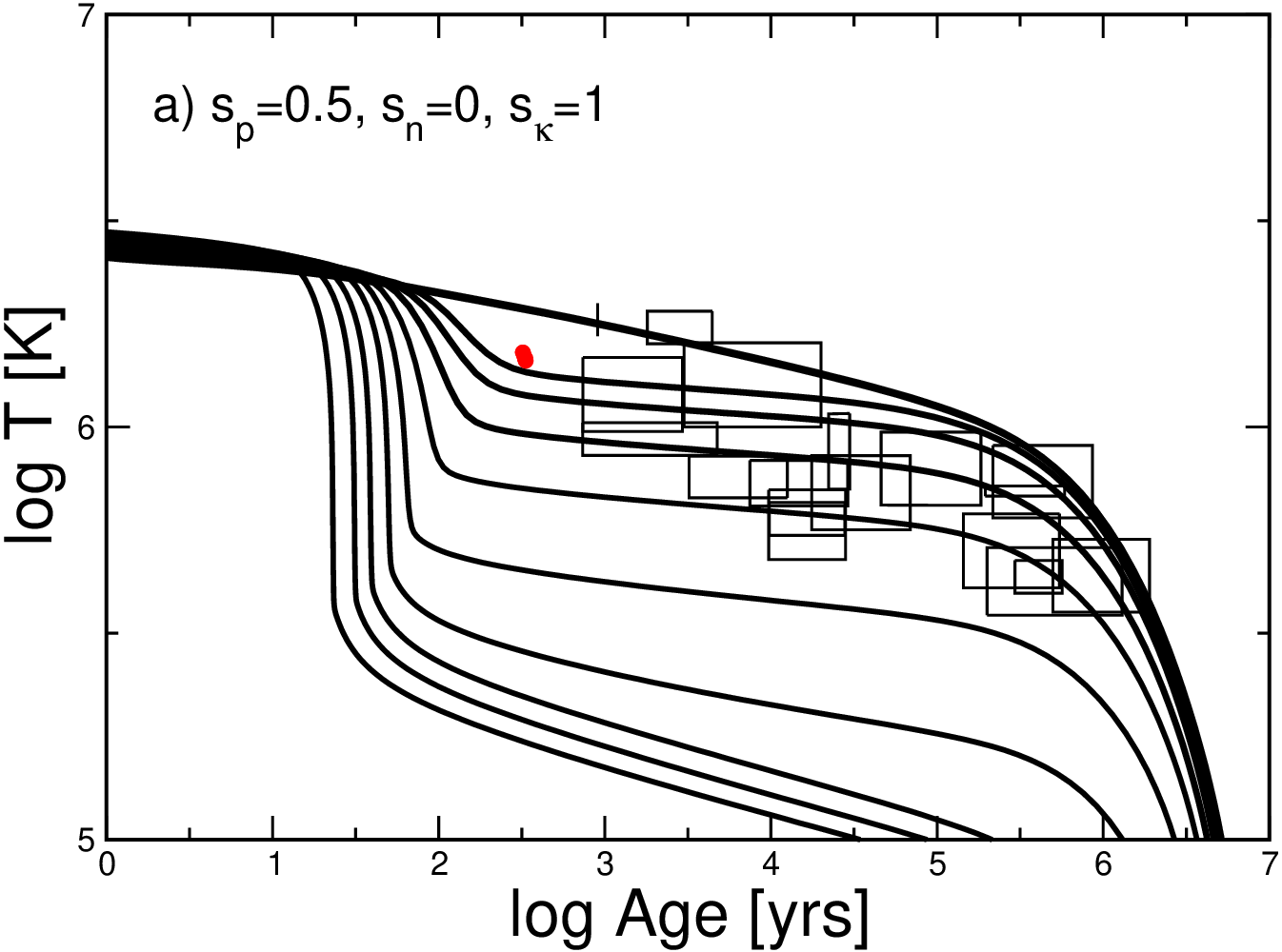}
\includegraphics[angle=0,scale=0.35,clip]{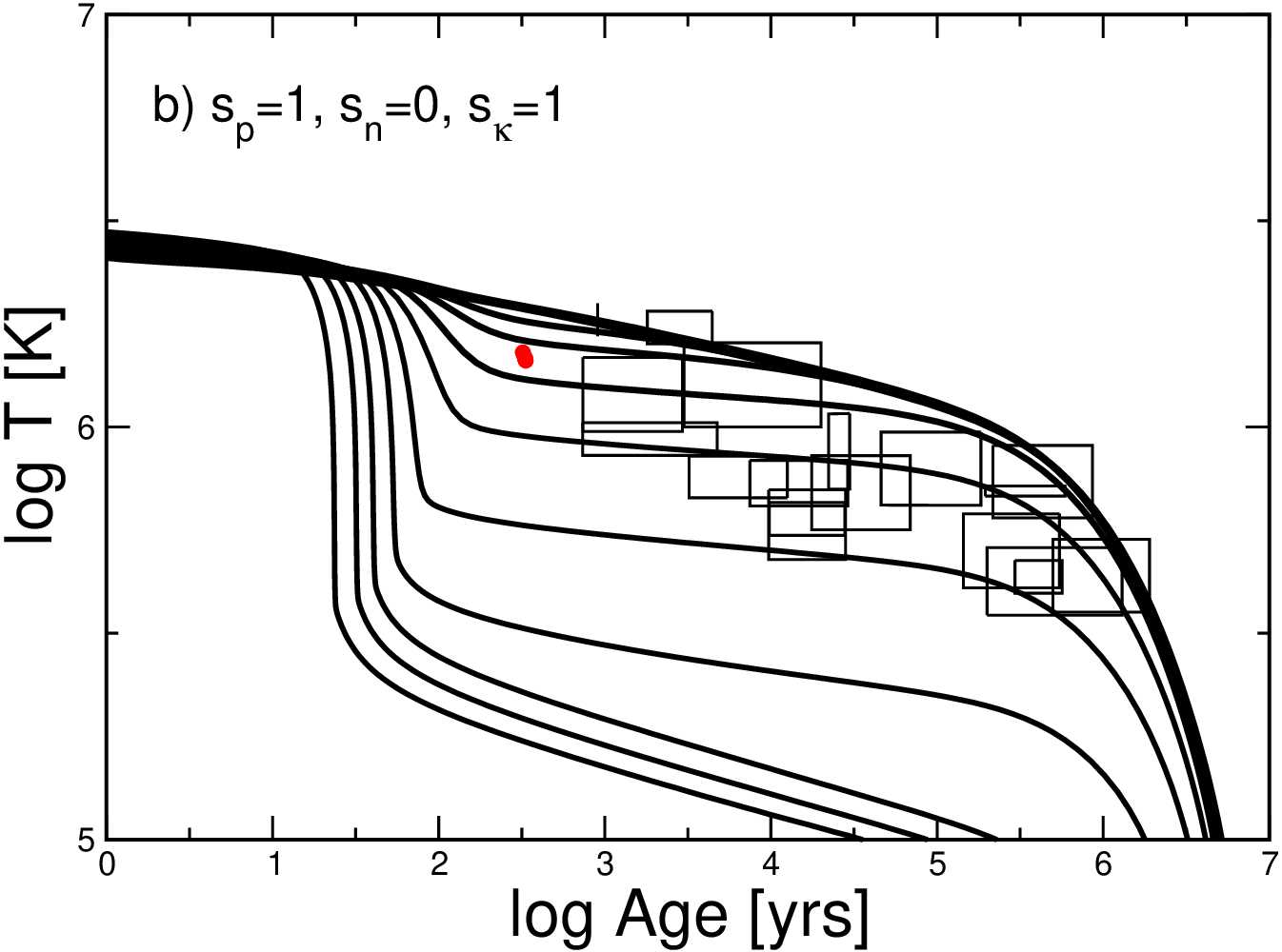}
\includegraphics[angle=0,scale=0.35,clip]{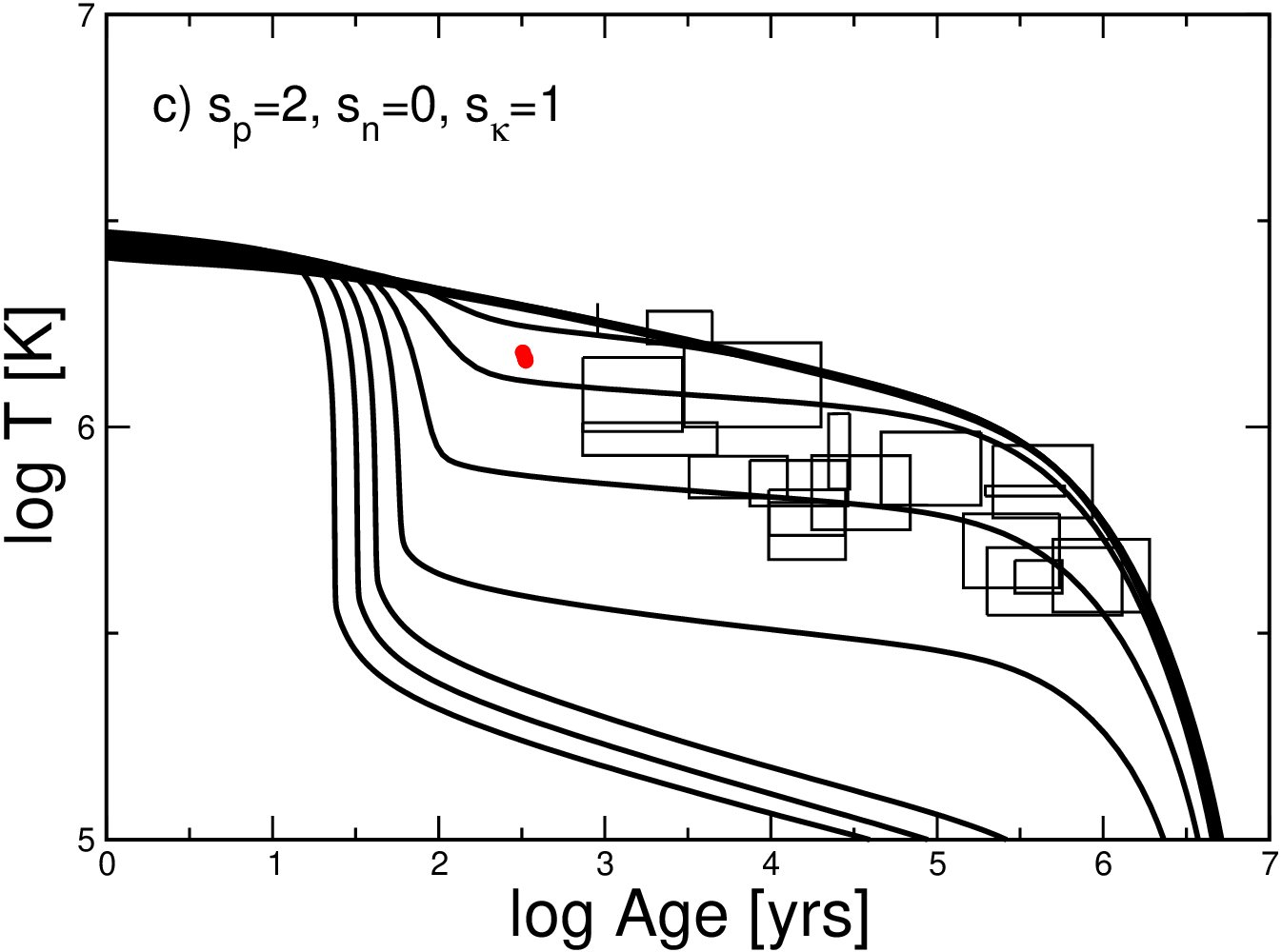}
\caption{(Color online)
Same as Fig.~\ref{f:bcs},
using $s_n=0$, $s_\kappa=1$, and different choices of
$s_p=0.5,1.0,2.0$ (from top to bottom).
}
\label{f:nocas}
\end{figure}

\subsection{Scenario 4: No Cas~A constraint}

The previous results have demonstrated that it is difficult to satisfy
simultaneously the rapid cooling of Cas~A and the slow cooling of old NS,
see Fig.~\ref{f:npb} or Fig.~\ref{f:del}(a),
where the hottest stars are slightly missed.
This is true not only in the current analysis \citep{pbf}.
However, recently doubts have been expressed about the validity of the
Cas~A data analysis \citep{casno},
such that a future revision towards much slower or no cooling at all
is not excluded.

We therefore study finally a scenario without the Cas~A constraint
(apart from reproducing its current age and temperature with a reasonable mass)
in our strong DU model,
trying to cover the full range of current cooling data.
Starting from the observation that the use of the unscaled BCS gaps in
Fig.~\ref{f:bcs} yields already a reasonable reproduction of most young NS,
and considering the fact that a finite n3P2 gap produces too strong PBF cooling,
simply switching off this channel
yields a nearly perfect coverage of all current cooling data,
as shown in Fig.~\ref{f:nocas}(b).
In this scenario Cas~A turns out a $1.31\ms$ NS.

Thus the BCS p1S0 gap alone is able to suppress sufficiently the DU cooling,
provided that it extends over a large enough density range.
Since in our case the p1S0 gap is perhaps somewhat large
(although this might be compensated by a different density shape),
we finally investigate the effect of rescaling it
with factors $s_p=0.5$ and $s_p=2$,
shown in panels (a) and (c).
It turns out that in both cases the quality of the cooling simulation
remains excellent,
just the predicted mass of Cas~A is varying between $1.18\ms$ and $1.46\ms$.
This illustrates the dire necessity of precise information on the masses
of the NS in the cooling diagram,
without which no theoretical cooling model can be verified.

A further possible constraint for cooling models could be the statistical
population synthesis $\log N - \log S$ analysis,
which might exclude a very sudden early onset of unmasked DU cooling
\citep{Popov2006,DUcon,BlaGri,PoPo1,PoPo2,Beznogov1,Beznogov2}.
A given cooling model,
together with the position of the observed NS in the cooling diagram,
makes a definite prediction for the mass spectrum of NS,
which could be compared with empirical information.
For example, a sudden onset of the DU process as in Fig.~\ref{f:del}(a)
causes a strong variation of the cooling curve with increasing NS mass,
and therefore a related enhancement of the deduced
NS mass spectrum in that domain.

However, this method is currently still burdened with large uncertainties,
in particular due to the unknown ``true" mass distribution of NS
and the difficult selection effect that makes too cold NS unobservable!
Furthermore, in our calculation the DU process is strongly masked by
the extended p1S0 gap,
such that the transition to DU cooling is not very abrupt
(see Fig.~\ref{f:nocas}),
and therefore an exclusion of this scenario might not be straightforward.
Together with a better knowledge of NS masses,
the population synthesis
might however be a further efficient tool for the cooling analysis
in the future.

\section{Conclusions}
\label{s:end}

We have studied NS cooling using a microscopic BHF EOS
featuring strong direct Urca reactions
setting in at $\rho=0.44\,\text{fm}^{-3}$, $M/\ms>1.10$,
and using compatible p1S0 and n3P2 pairing gaps
as well as nucleon effective masses.
The current substantial theoretical uncertainty regarding gaps and
thermal conductivity was modelled in a rather simple way by introducing
three global scale factors.

We found that it is possible to reproduce the apparent fast cooling of
Cas~A by either finetuning the n3P2 pairing gap or reducing the
thermal conductivity.
In general it is difficult to then simultaneously fit old hot NS,
although we did find a suitable parameter set for that purpose.

Relaxing the Cas~A constraint,
it is astonishing to see how well all current cooling data can be fit
by just assuming the p1S0 BCS gap
(with some freedom of scaling)
and a vanishing n3P2 gap.

Our results affirm the extreme difficulty to draw quantitative conclusions
from the current NS cooling data containing no information on the masses
of the cooling objects,
due to the large variety of required microphysics input
that is hardly known or constrained otherwise.

We have shown that not even the combination of very strong DU cooling
with sufficiently large and extended p1S0 gaps
and small n3P2 pairing gaps can be excluded.
There is still ample freedom to choose the magnitude and shape of the p1S0 gap,
as long as the covered density domain is sufficiently large
in order to fully mask the DU onset up to sufficiently heavy stars.
One can only hope to resolve this problem
once precise information on the NS masses in the cooling diagram
becomes available.

Even more exotic possibilities of blocking the DU process
by strong p3P2 pairing \citep{ourgaps} are not excluded either,
but were not analyzed in this work;
as neither the effect of exotic components of matter (hyperons, quarks)
that should appear at high density
and completely change the theoretical picture \citep{exo1,exo2}.
In any case there are strong indications from theoretical many-body calculations
and supported by the current analysis,
that the DU process becomes active at moderately high baryon density;
it should thus never be excluded without justification in cooling simulations.

\section*{Acknowledgments}

We acknowledge useful discussions with M.~Baldo, A.~Bonanno, D.~Page, and W. Ho.
Partial support comes from ``NewCompStar," COST Action MP1304.

\bibliographystyle{mnras}     
\bibliography{cooldu}         

\bsp                          
\label{lastpage}
\end{document}